\documentclass{article}
\usepackage{authblk}
\usepackage[table]{xcolor}
\usepackage[colorlinks]{hyperref}
\usepackage{natbib}
\usepackage{enumerate}
\usepackage[margin=1in]{geometry}
\usepackage{subcaption}
\usepackage{amsmath}
\usepackage{microtype}

\usepackage{tikz}
\usetikzlibrary{arrows, automata,positioning}

\definecolor{midnightblue}{rgb}{0.1, 0.1, 0.44}
\hypersetup{
    pdfauthor={Qirun Zhang},
    plainpages=false,
    linkcolor=black, 
    citecolor=midnightblue, 
    filecolor=black,
    urlcolor=midnightblue,
    pdfpagelabels
}
\date{}

\newcommand{\codeIn}[1]{{\small\tt{#1}}}
\title{A Note on Dynamic Bidirected Dyck-Reachability with  Cycles}
\author{Qirun Zhang}
\affil{School of Computer Science, Georgia Institute of Technology}
\affil{\url{qrzhang@gatech.edu}}
\begin{document}
\maketitle
\begin{abstract}
Recently, \citet{LiSZ22} presented a dynamic Dyck-reachability algorithm for bidirected graphs. The basic idea is based on updating edge weights in a data structure called the \emph{merged graph} $G_m$. As noted in~\citet{popl}, the edge deletion procedure described in the algorithm of \citet{LiSZ22} cannot properly update the weights in the presence of cycles in $G_m$. This note discusses the cycle case and the time complexity.
\end{abstract}

\section{Merged Graph $G_m$}

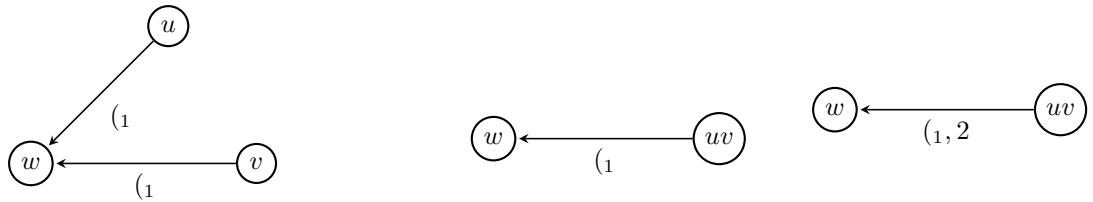
\begin{figure}
\centering
\begin{subfigure}{0.45\textwidth}\centering
\begin{tikzpicture}[
            > = stealth, 
            shorten > = 1pt, 
            auto,
            node distance = 3cm, 
            semithick 
        ]

        \tikzstyle{every state}=[
            draw = black,
            thick,
            fill = white, inner sep=1mm,
            minimum size = 3mm
        ]

        \node[state] (s) {$w$};
        \node[state] (v1) [above right  = 2cm of s] {$u$};
        \node[state] (v2) [right of=s] {$v$};

        \path[->] (v1) edge node {$(_1$} (s);
        \path[->] (v2) edge node {$(_1$} (s);

    \end{tikzpicture}
    \caption{Input bidirected graph $G$. Note that for each open-parenthesis-labeled  edge $u\xrightarrow{(_i}v$, there exists a close-parenthesis-labeled edge $v\xrightarrow{)_i}u$,
and vice versa. We omit close-parenthesis-labeled edges for brevity.}\label{fig:dyck1}
    \end{subfigure}\hfill\begin{subfigure}{0.25\textwidth}\centering
\begin{tikzpicture}[
            > = stealth, 
            shorten > = 1pt, 
            auto,
            node distance = 3cm, 
            semithick 
        ]

        \tikzstyle{every state}=[
            draw = black,
            thick,
            fill = white, inner sep=1mm,
            minimum size = 3mm
        ]

        \node[state] (s) {$w$};
        \node[state] (v2) [right of=s] {$uv$};
            \node[] (s1) [below =1cm of s]  {};
        
        \path[->] (v2) edge node {$(_1$} (s);

    \end{tikzpicture}
       \caption{Merged graph $G_m$ used in bidirected Dyck-reachability algorithms.}\label{fig:dyck2}
    \end{subfigure}\hfill\begin{subfigure}{0.25\textwidth}\centering
\begin{tikzpicture}[
            > = stealth, 
            shorten > = 1pt, 
            auto,
            node distance = 3cm, 
            semithick 
        ]

        \tikzstyle{every state}=[
            draw = black,
            thick,
            fill = white, inner sep=1mm,
            minimum size = 3mm
        ]

        \node[state] (s) {$w$};
        \node[state] (v2) [right of=s] {$uv$};
               \node[] (s1) [below =1cm of s]  {};
        
        \path[->] (v2) edge node {$(_1, 2$} (s);

    \end{tikzpicture}
    \caption{Weighted merged graph $G_m$ used in \citet{LiSZ22}'s dynamic bidirected Dyck-reachability algorithm.}\label{fig:dyck3}
    \end{subfigure}
    \caption{Bidirected Dyck-reachability.}\label{fig:dyck}
\end{figure}

The algorithm proposed by \citet{LiSZ22} utilizes a core data structure called the merged graph $G_m$. This data structure realizes the equivalence relation for bidirected Dyck-reachability~\citep{ZhangLYS13,ChatterjeeCP18}. Consider an input (bidirected) graph $G$, if two nodes $u$ and $v$ in $G$ are Dyck-reachable, we can merge them and obtain a more compact graph representation $G_m$. Each node in $G_m$ represents a set of Dyck-reachable nodes in $G$, and each edge represents the corresponding merged edges in $G$.  Consider the example in Figure~\ref{fig:dyck}. Figure~\ref{fig:dyck1} gives the input bidirected graph, and Figure~\ref{fig:dyck2} gives the corresponding merged graph $G_m$. In this example, nodes $u$ and $v$ are Dyck-reachable in $G$, and they are represented by the same representative node $uv$ in $G_m$.

\section{Cycles in Merged Graphs}\label{sec:cycle}
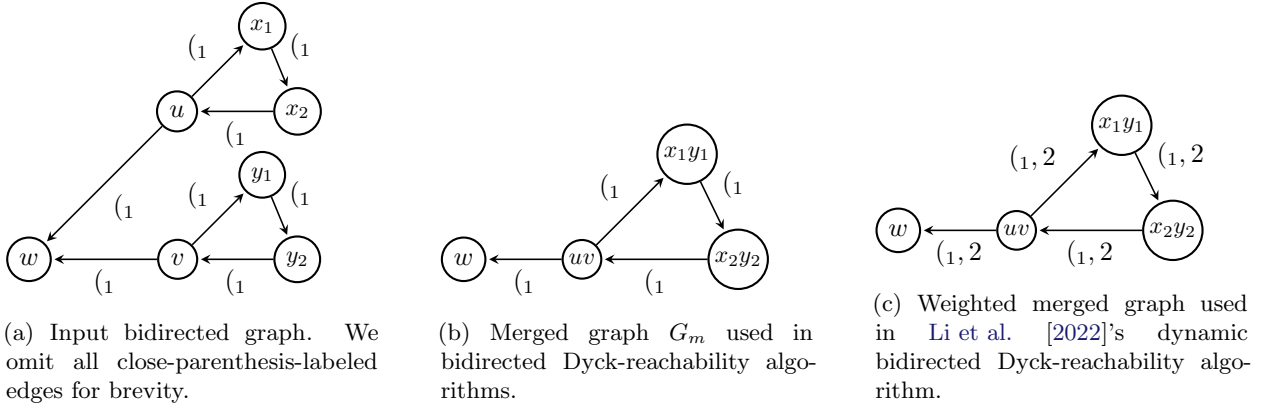
\begin{figure}
\centering
\begin{subfigure}{0.3\textwidth}
\begin{tikzpicture}[
            > = stealth, 
            shorten > = 1pt, 
            auto,
            node distance = 3cm, 
            semithick 
        ]

        \tikzstyle{every state}=[
            draw = black,
            thick,
            fill = white, inner sep=1mm,
            minimum size = 3mm
        ]

        \node[state] (s) {$w$};
        \node[state] (v1) [above right  = 2.2cm of s] {$u$};
        \node[state,scale=.9] (x1) [above right  = 1cm of v1] {$x_1$};
        \node[state,scale=.9] (x2) [right  = 1cm of v1] {$x_2$};
        \node[state] (v2) [right = 1.4cm of s] {$v$};
        \node[state,scale=.9] (y1) [above right  = 1cm of v2] {$y_1$};
        \node[state,scale=.9] (y2) [right  = 1cm of v2] {$y_2$};
   
        \path[->] (v1) edge node {$(_1$} (s);
        \path[->] (v2) edge node {$(_1$} (s);
        \path[->] (v2) edge node {$(_1$} (y1);
        \path[->] (y1) edge node {$(_1$} (y2);
        \path[->] (y2) edge node {$(_1$} (v2);
        \path[->] (v1) edge node {$(_1$} (x1);
        \path[->] (x1) edge node {$(_1$} (x2);
        \path[->] (x2) edge node {$(_1$} (v1);

    \end{tikzpicture}
    \caption{Input bidirected graph. We omit all close-parenthesis-labeled edges for brevity.}\label{fig:dyndyck1}
    \end{subfigure}\hfill\begin{subfigure}{0.3\textwidth}
\begin{tikzpicture}[
            > = stealth, 
            shorten > = 1pt, 
            auto,
            node distance = 3cm, 
            semithick 
        ]

        \tikzstyle{every state}=[
            draw = black,
            thick,
            fill = white, inner sep=1mm,
            minimum size = 3mm
        ]

        \node[state] (s) {$w$};
             \node[state,scale=.9,inner sep=.6mm] (v2) [right = 1cm of s] {$uv$};
        \node[state,scale=.9,inner sep=.6mm] (y1) [above right  = 1.3cm of v2] {$x_1y_1$};
        \node[state,scale=.9,inner sep=.6mm] (y2) [right  = 1.4cm of v2] {$x_2y_2$};

        \path[->] (v2) edge node {$(_1$} (s);
        \path[->] (v2) edge node {$(_1$} (y1);
        \path[->] (y1) edge node {$(_1$} (y2);
        \path[->] (y2) edge node {$(_1$} (v2);

    \end{tikzpicture}
    \caption{Merged graph $G_m$ used in
bidirected Dyck-reachability
algorithms.}\label{fig:dyndyck2}
    \end{subfigure}\hfill\begin{subfigure}{0.3\textwidth}
\begin{tikzpicture}[
            > = stealth, 
            shorten > = 1pt, 
            auto,
            node distance = 3cm, 
            semithick 
        ]

        \tikzstyle{every state}=[
            draw = black,
            thick,
            fill = white, inner sep=1mm,
            minimum size = 3mm
        ]

        \node[state] (s) {$w$};
             \node[state,scale=.9,inner sep=.6mm] (v2) [right = 1cm of s] {$uv$};
        \node[state,scale=.9,inner sep=.6mm] (y1) [above right  = 1.3cm of v2] {$x_1y_1$};
        \node[state,scale=.9,inner sep=.6mm] (y2) [right  = 1.4cm of v2] {$x_2y_2$};

        \path[->] (v2) edge node {$(_1,2$} (s);
        \path[->] (v2) edge node {$(_1,2$} (y1);
        \path[->] (y1) edge node {$(_1,2$} (y2);
        \path[->] (y2) edge node {$(_1,2$} (v2);

    \end{tikzpicture}
    \caption{Weighted merged graph used in \citet{LiSZ22}'s dynamic bidirected Dyck-reachability algorithm.}\label{fig:dyndyck3}
    \end{subfigure}
    \caption{Bidirected Dyck-reachability with cycles. }\label{fig:dyndyck}
\end{figure}

Cycles may exist in the merged graph $G_m$. Figure~\ref{fig:dyndyck} gives such an example. This example is a slightly modified version of Figure 10 in \citet{popl}, where the cycles have been extended.

The distinction between the acyclic case (Figure~\ref{fig:dyck}) and the cyclic case (Figure~\ref{fig:dyndyck}) is that, in the acyclic case, nodes $u$ and $v$ are merged due to an anchor node $w$. However, in the cyclic case, upon observing the merged graph $G_m$ in Figure~\ref{fig:dyndyck2} alone, there are two anchor nodes $w$ and $x_1y_1$, where $x_1y_1$ depends on $w$. Specifically, with reference to the input graph in Figure~\ref{fig:dyndyck1}, it is evident that the merging of $u$ and $v$ is a result of node $w$. The merging of nodes $x_1$ and $y_1$ is due to $w$ as well, which subsequently leads to the potential merging of $n$ and $v$ where they have already been merged. Therefore, the merging of $u$ and $v$ does not depend on the merging of $x_1$ and $y_1$. In the merged graph $G_m$ in Figure~\ref{fig:dyndyck2}, we can see that the nodes $x_1y_1$ and $uv$ are in a cycle. Consequently, it appears that the merged node $x_1y_1$ is responsible for the merging of $u$ and $v$.

\section{Dynamic Bidirected Dyck-Reachability Algorithm}\label{sec:algo}
The dynamic algorithm proposed by \citet{LiSZ22} relies on the merged graph $G_m$. The high-level idea is to assign weights to each edge in $G_m$ based on the number of merged edges in the original graph $G$. In Figures~\ref{fig:dyck3} and~\ref{fig:dyndyck3}, we can see examples of weighted merged graphs, where each edge represents two merged edges from the corresponding input graphs in Figures~\ref{fig:dyck1} and~\ref{fig:dyndyck1}, respectively. Upon insertion or deletion of an edge, the weights are updated accordingly. If the weight of an outgoing edge from a representative node $n'$ in $G_m$ changes to $1$, it means that the input graph nodes represented by $n'$ may no longer be Dyck-reachable in $G$. In such cases, we need to potentially split the representative node $n'$.  This approach works well when $G_m$ is acyclic. 

If the merged graph $G_m$ contains cycles, it becomes challenging to track the anchor nodes. For example, in the weighted merged graph $G_m$ shown in Figure~\ref{fig:dyndyck3},  nodes $w$ and $x_1y_1$  appear to be the anchor nodes that create the representative node $uv$ in $G_m$ based solely on their weights.  As a result, they become indistinguishable. However, according to the discussion in Section~\ref{sec:cycle}, node $uv$ is created before the creation of $x_1y_1$ ,and node $x_1y_1$ actually depends on $w$.  

\begin{figure}
\centering
\begin{subfigure}{0.3\textwidth}
\begin{tikzpicture}[
            > = stealth, 
            shorten > = 1pt, 
            auto,
            node distance = 3cm, 
            semithick 
        ]

        \tikzstyle{every state}=[
            draw = black,
            thick,
            fill = white, inner sep=1mm,
            minimum size = 3mm
        ]

        \node[state] (s) {$w$};
        \node[state] (v1) [above right  = 2.2cm of s] {$u$};
        \node[state,scale=.9] (x1) [above right  = 1cm of v1] {$x_1$};
        \node[state,scale=.9] (x2) [right  = 1cm of v1] {$x_2$};
        \node[state] (v2) [right = 1.4cm of s] {$v$};
        \node[state,scale=.9] (y1) [above right  = 1cm of v2] {$y_1$};
        \node[state,scale=.9] (y2) [right  = 1cm of v2] {$y_2$};
   
        \path[->] (v1) edge node {$(_1$} (s);
        \path[->] (v2) edge node {$(_1$} (y1);
        \path[->] (y1) edge node {$(_1$} (y2);
        \path[->] (y2) edge node {$(_1$} (v2);
        \path[->] (v1) edge node {$(_1$} (x1);
        \path[->] (x1) edge node {$(_1$} (x2);
        \path[->] (x2) edge node {$(_1$} (v1);

    \end{tikzpicture}
    \caption{Bidirected graph after removing $v\xrightarrow{(_1}w$.}\label{fig:del1}
    \end{subfigure}\hfill\begin{subfigure}{0.3\textwidth}
\begin{tikzpicture}[
            > = stealth, 
            shorten > = 1pt, 
            auto,
            node distance = 3cm, 
            semithick 
        ]

        \tikzstyle{every state}=[
            draw = black,
            thick,
            fill = white, inner sep=1mm,
            minimum size = 3mm
        ]

        \node[state] (s) {$w$};
             \node[state,scale=.9,inner sep=.6mm] (v2) [right = 1cm of s] {$uv$};
        \node[state,scale=.9,inner sep=.6mm] (y1) [above right  = 1.3cm of v2] {$x_1y_1$};
        \node[state,scale=.9,inner sep=.6mm] (y2) [right  = 1.4cm of v2] {$x_2y_2$};

        \path[->] (v2) edge node {$(_1,1$} (s);
        \path[->] (v2) edge node {$(_1,2$} (y1);
        \path[->] (y1) edge node {$(_1,2$} (y2);
        \path[->] (y2) edge node {$(_1,2$} (v2);

    \end{tikzpicture}
    \caption{Weighted merged graph $G_m$ computed by \codeIn{DynamicDeletion} in \citet{LiSZ22}.}\label{fig:del2}
    \end{subfigure}\hfill\begin{subfigure}{0.3\textwidth}
\begin{tikzpicture}[
            > = stealth, 
            shorten > = 1pt, 
            auto,
            node distance = 3cm, 
            semithick 
        ]

        \tikzstyle{every state}=[
            draw = black,
            thick,
            fill = white, inner sep=1mm,
            minimum size = 3mm
        ]

        \node[state] (s) {$w$};
        \node[state] (v1) [above right  = 2.2cm of s] {$u$};
        \node[state,scale=.9] (x1) [above right  = 1cm of v1] {$x_1$};
        \node[state,scale=.9] (x2) [right  = 1cm of v1] {$x_2$};
        \node[state] (v2) [right = 1.4cm of s] {$v$};
        \node[state,scale=.9] (y1) [above right  = 1cm of v2] {$y_1$};
        \node[state,scale=.9] (y2) [right  = 1cm of v2] {$y_2$};
   
        \path[->] (v1) edge node[above, midway] {$(_1,1$} (s);
        \path[->] (v2) edge node {$(_1,1$} (y1);
        \path[->] (y1) edge node {$(_1,1$} (y2);
        \path[->] (y2) edge node {$(_1,1$} (v2);
        \path[->] (v1) edge node {$(_1,1$} (x1);
        \path[->] (x1) edge node {$(_1,1$} (x2);
        \path[->] (x2) edge node {$(_1,1$} (v1);

    \end{tikzpicture}
    \caption{Weighted merged graph $G_m$ computed by the extended algorithm in Section~\ref{sec:algo}.}\label{fig:del3}
    \end{subfigure}
    \caption{Handling bidirected Dyck-reachability with cycles. }\label{fig:del}
\end{figure}
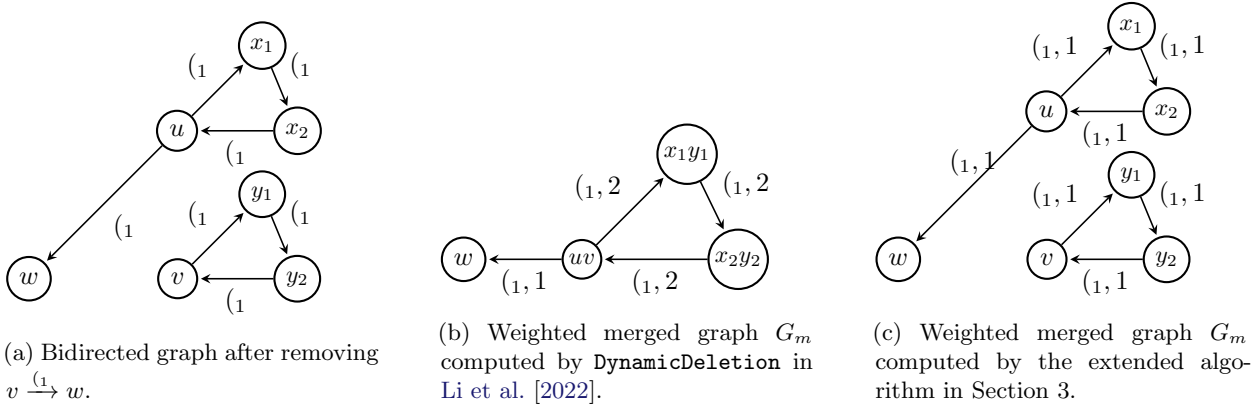

Now, consider the removal of the edge $v\xrightarrow{(_1}w$ in Figure~\ref{fig:dyndyck1}. Figure~\ref{fig:del1} depicts the input graph after the edge removal. As noted in \citet{popl}, after removing this edge in $G$, the deletion algorithm given in \citet{LiSZ22} cannot properly split the node $uv$ because it is still connected to a spurious anchor node $x_1y_1$ in $G_m$ (Figure~\ref{fig:del2}). The root cause of this problem is that the weight condition (Definition 4.2) given in \citet{LiSZ22} only holds when a merging node $x$ in $G_m$ is formed after the formation of its corresponding anchor node $y$, meaning that $x$ depends on $y$. In the presence of cycles, as illustrated in Figure~\ref{fig:dyndyck2}, nodes $uv$ and $x_1y_1$ are merged simultaneously, indicating that they are mutually dependent and both nodes depend on $w$.

The dynamic algorithm (Algorithm 1) in \citet{LiSZ22} should be extended to handle cycles. The insertion algorithm remains the same. For edge deletion, we should apply the following steps.
Assume the removal of an edge $e=u\xrightarrow{(_i} v$ in $G$.
\begin{enumerate}[(i):]
    \item \emph{Tracking cycles.} Perform a DFS from the representative node \codeIn{resp\_node(u)} in $G_m$ by traversing only the incoming edges. Then, collect all nodes involved in a cycle and store them in a set $N_c$. 
    \item \emph{Splitting nodes.} If $N_c == \emptyset$, call the original \codeIn{DynamicDeletion} procedure (\emph{i.e.}, Algorithm 3 in \citet{LiSZ22}). Otherwise, for each $n \in N_c$, split $n$ in the merged graph $G_m$, and run \codeIn{Opt-Dyck'}~\citep{ChatterjeeCP18,LiSZ22} on $G_m$.
\end{enumerate}

Recall the example of removing $v\xrightarrow{(_1}w$ in Figure~\ref{fig:dyndyck1}. The extended algorithm traverses the nodes in $G_m$ (Figure~\ref{fig:dyndyck2}) and creates the set $N_c = \{uv, x_2y_2, x_1y_1\}$. Then, it splits the nodes in $N_c$ and generates a graph that is identical to the input graph shown in Figure~\ref{fig:del1}. After running the
 \codeIn{Opt-Dyck'} algorithm, we obtain the weighted merged graph as shown in Figure~\ref{fig:del3}.

\section{Discussion}
The extended algorithm presented in Section~\ref{sec:algo} splits all strongly-connected-component (SCC) nodes in the merged graph $G_m$ and applies the optimal static Dyck-reachability algorithm \codeIn{Opt-Dyck'}~\citep{ChatterjeeCP18,LiSZ22}. The correctness follows \codeIn{Opt-Dyck'}.  If the merged graph $G_m$ does not contain any cycles, the \codeIn{DynamicDeletion} procedure correctly handles edge deletions~\citep{LiSZ22}. Among all the operations in this algorithm, the most expensive step is the edge merging associated with node splitting, even in the acyclic case. The worst-case complexity of the algorithm by \citet{LiSZ22} is quadratic~\citep{popl}.  Consider lines 4-6 in the \codeIn{split\_further} procedure (Procedure 4 in \citet{LiSZ22}). In the worst-case scenario, a particular node in $G_m$ may have $O(n)$ outgoing edges, where each of these edges could represent $O(n)$ original edges in the input graph $G$. The complexity analysis of lines 12-18 in procedure \codeIn{split\_further} and step (ii) in our extended algorithm (Section~\ref{sec:algo}) is similar. Therefore, the overall time complexity of the extended algorithm for dynamic Dyck-reachability is $O(n^2)$.

Handling cycles in the merged graph $G_m$ is nontrivial. One possible improvement to our extended algorithm in Section~\ref{sec:algo} is to eliminate the DFS in step (i) by dynamically maintaining the topological ordering in $G_m$~\citep{DBLP:journals/talg/BenderFGT16}. To improve the $O(n^2)$ time complexity, it is crucial to effectively handle node splitting. Notably, a recent result by \citet{popl} maintains Dyck SCCs through multiple layers of abstractions and handles edge deletion in $O(n\cdot \alpha(n))$ time.

\subsection*{Acknowledgments}
We would like to thank Andreas Pavlogiannis for communicating the results, as well as the POPL'24 organizers for their assistance.

\bibliographystyle{plainnat} 
\bibliography{ref} 
\end{document}